\documentclass[prl,aps,twocolumn,superscriptaddress,preprintnumbers,floatfix,longbibliography]{revtex4-1}
\usepackage{amsmath,amssymb,amsfonts,bm,graphicx,epsf,colordvi,bbm,pifont,siunitx,enumitem,hyperref,xcolor,multirow}
\usepackage[caption=false]{subfig}
\pdfoutput=1

\begin{document} 
\title{Anomaly Detection for Resonant New Physics with Machine Learning}
%\title{Model Independent Searches for Physics Beyond the Standard Model with Machine Learning}
\author{Jack Collins}
\email{jhc296@umd.edu}
\affiliation{Maryland Center for Fundamental Physics, Department of Physics, University of Maryland, College Park, MD 20742, USA}
\affiliation{Department of Physics and Astronomy, Johns Hopkins University, Baltimore, MD 21218, USA}

\author{Kiel Howe}
\email{khowe@fnal.gov}
\affiliation{Fermi National Accelerator Laboratory, Batavia IL 60510, USA}

\author{Benjamin Nachman}
\email{bpnachman@lbl.gov}
\affiliation{Physics Division, Lawrence Berkeley National Laboratory, Berkeley, CA 94720, USA}
\affiliation{Simons Institute for the Theory of Computing, University of California, Berkeley, Berkeley, CA 94720, USA}

\newcommand{\jc}[1]{{\color{magenta} #1}}
\newcommand{\bn}[1]{{\color{blue} #1}}

\begin{abstract}
Despite extensive theoretical motivation for physics beyond the Standard Model (BSM) of particle physics, searches at the Large Hadron Collider (LHC) have found no significant evidence for BSM physics.  Therefore, it is essential to broaden the sensitivity of the search program to include unexpected scenarios.  We present a new model-agnostic anomaly detection technique that naturally benefits from modern machine learning algorithms.  The only requirement on the signal for this new procedure is that it is localized in at least one known direction in phase space.  Any other directions of phase space that are uncorrelated with the localized one can be used to search for unexpected features.  This new method is applied to the dijet resonance search to show that it can turn a modest $2\sigma$ excess into a $7\sigma$ excess for a model with an intermediate BSM particle that is not currently targeted by a dedicated search.
\end{abstract}

\maketitle

The main goal of high energy physics is to identify the elementary building blocks of matter and to characterize the laws governing their motion.  In order to achieve this goal, experiments at the energy frontier collide particles with extremely high momenta in a quest to directly produce the elementary particles and study their interactions.  The collider currently able to directly probe the smallest distance scales is the Large Hadron Collider (LHC).  Building on decades of effort at previous experiments, the ATLAS and CMS collaborations at the LHC discovered the Higgs boson in 2012~\cite{Aad:2012tfa,Chatrchyan:2012xdj}, completing the Standard Model (SM) of particle physics.  While the SM has been enormously successful, it is not a complete theory of nature as it lacks a description of dark matter and gravity, in addition to various technical or aesthetic problems.  Despite an intensive and impressive program to search directly for physics beyond the SM at the LHC~\cite{atlassusytwiki,atlasexoticstwiki,cmsexoticstwiki,cmssusytwiki,cmsb2gtwiki}, there is still no direct evidence for any new structures in nature.  However, there are numerous compelling theoretical motivations for physics beyond the SM (BSM) at the energies scales accessible by the LHC~\cite{Ellis:2007zzc}.  While it could be that the BSM particles are too massive or produced with too low a cross-section to be discovered yet, it is also possible that the current search program is simply not sensitive to the regions of phase space populated by BSM physics.  

In order to mitigate the possibility of uncovered regions of phase space, collider experiments have implemented model-independent anomaly detection techniques.  Traditionally, there are two such approaches: general searches and bump hunts.  The idea of general searches is to compare data and simulation in a large number of event topologies, characterized by the number and type of various physics objects such as leptons or hadronic jets resulting from high energy quark and gluon production~\cite{ATLAS-CONF-2017-001,CMS-PAS-EXO-10-021,ATLAS-CONF-2012-107,ATLAS-CONF-2014-006,Aktas:2004pz,Aaron:2008aa,Abbott:2000fb,Abbott:2000gx,Aaltonen:2007dg,Aaltonen:2008vt,sleuth,Knuteson:2004nj}.  While this approach has a broad coverage, it is restricted to simple observables because it relies heavily on simulations for background estimation.  In contrast, bump hunts~\cite{Choudalakis:2011qn} often do not use any simulation for background estimation, other than to motivate and validate the background fit procedure: after identifying a region of phase space where a signal is expected to be localized, the background is fit with a smooth function and interpolated to the signal-sensitive region.  Excesses over this background prediction would be an indication of BSM physics.  To enhance the resonance structure from a di-object invariant mass, modern classification tools~\footnote{The first uses of jet substructure for particle searches were proposed in Ref.~\cite{Seymour:1993mx,Butterworth:2002tt,Butterworth:2008iy}; for recent theoretical and experimental reviews, see Ref.~\cite{Larkoski:2017jix} and Ref.~\cite{Asquith:2018igt}, respectively.} can be used select the target objects like $b$-quark~\cite{ATL-PHYS-PUB-2017-013,CMS-PAS-BTV-15-001}, top-quark~\cite{ATL-PHYS-PUB-2017-004,CMS-DP-2017-026}, $W/Z$~\cite{ATL-PHYS-PUB-2017-004,CMS-DP-2017-026} or Higgs boson~\cite{ATLAS-CONF-2016-039,CMS-PAS-BTV-15-002} jets from generic quark or gluon jets.  However, these classifiers are trained in simulation and calibrated in data, which may lead to suboptimal classifiers.   Furthermore, it is not possible in this paradigm to develop classifiers for BSM objects, since no calibration sample exists.

This letter presents a new technique to search for BSM physics that significantly extends the bump hunt approach that uses classifiers trained directly on data.  Consider a signal that is localized in one kinematic variable (the resonant variable, $m_\text{res}$) on top of a smoothly varying background, for example a dijet resonance that can be reconstructed from the invariant mass of two jets.  Suppose that each event has additional auxiliary information (such as substructure in the two jets) that may provide additional discriminating power between signal and background, but the detailed signal characteristics in these auxiliary variables are unknown a priori.  Our proposal is that a classifier can be trained to discern the auxiliary characteristics of the signal (if present) directly from data, without reference to any specific signal model hypothesis.  The output of this classifier can then be used to select signal-like events and reject background events, producing a new distribution in the resonant variable that remains smooth in the case that no signal was present, but that may enhance the significance of the bump if a real signal is present.  In the event that a signal is discovered, the output of the classifier can then be studied to infer the signal characteristics.

The key feature of resonant signals that is utilized in our approach is that their localization in one kinematic variable on top of a smoothly varying background allows the identification of potential signal-enhanced and signal-depleted signal and sideband regions, respectively, with almost identical background characteristics.  A classifier trained to distinguish the auxiliary characteristics of the signal region events from those of the sideband may in principle be as powerful as a classifier trained to distinguish pure samples of signal and background events -- this is a specific application of Classification Without Labels (CWoLa)~\cite{Metodiev:2017vrx}.   To see why this is the case, suppose that it is possible to define an ideal sideband selection that contains only background and no signal, and an ideal signal region that contains background identical to that in the sideband but also a small signal that is distinct from the background.  By the Neyman--Pearson lemma~\cite{Neyman289}, the most powerful test statistic for discriminating signal (sig) events from background (bg) events using some observables $Y$ is the likelihood ratio
\begin{equation}
\label{equ:fullsupLikelihood}
L\left(Y\right) = \frac{p(Y | \mathrm{sig})}{p(Y| \mathrm{bg})},
\end{equation}
and a fully supervised classifier is trained to approximate any monotonic rescaling of this function.  A classifier that is trained to discriminate signal region events (sig+bg) from sideband region events (bg) will instead ideally learn to approximate a monotonic rescaling of the function
\begin{equation}
\label{equ:fullsupLikelihood2}
\hat{L}\left(Y\right) = \frac{p(Y | \mathrm{sig + bg})}{p(Y | \mathrm{bg})} = f_\text{sig} \frac{p(Y | \mathrm{sig})}{p(Y| \mathrm{bg})} + f_\text{bg},
\end{equation}
where $f_\text{sig}$ and $f_\text{bg}$ are the proportions of signal and background events in the signal region.  The fact that Eq.~\ref{equ:fullsupLikelihood2} is itself a monotonic rescaling of $L\left(Y\right)$ from Eq.~\ref{equ:fullsupLikelihood} shows that there is no fundamental obstruction for the CWoLa-based classifier to identify the ideal decision boundaries for signal selection.   The above argument also holds if the sideband region has a small amount of signal, as long as the signal proportion is less than the signal region~\cite{Metodiev:2017vrx}.  Practical limits will arise from limited statistics (particularly for the signal) and other technical difficulties that may obstruct a trainable classifier from reaching the performance achievable with labeled simulations, and also from the small differences in the background characteristics between signal and sideband regions.

Our extended bump hunt procedure also has some features in common with the sPlot~\cite{Pivk:2004ty} technique.  In particular, sPlot provides a method for determining the distribution of multiple event classes for a resonant feature (`control variable' in the language of Ref.~\cite{Pivk:2004ty}) using a set of uncorrelated auxiliary features (`discriminating variables' in Ref.~\cite{Pivk:2004ty}).  The key differences between sPlot and the extended bump hunt are (1) we are interested in using machine learning to isolate a signal-rich region of phase space and (2) we do not take the probability distribution for the auxiliary features as input - the classification procedure learns useful information directly from the data.

A danger that is present when training and testing a classifier on the same dataset is that it may overfit the training data and learn the specific statistical fluctuations in that dataset rather than the true underlying distribution.  Classifiers used in this way will preferentially select signal-region events based on their statistical fluctuations, and will create a fake bump in the resonance-variable distribution even when no real signal is present. A simple way to mitigate the background sculpting is to split the underlying dataset randomly into a training set and a test set that will have uncorrelated statistical fluctuations.  This would, however, result in an effective loss of luminosity available both for training and for testing. Instead we advocate for an $n$-fold cross-validation procedure, in which the data is randomly partitioned into $n$ sets of equal size (stratified by $m_{\text{res}}$ bin). The selection on each of the $n$ partitions is performed using the output of a classifier trained and validated on the remaining $n-1$ partitions, resulting in a total of $n$ classifiers.  Any statistical fluctuations learnt by a classifier from its training data will be uncorrelated with those in the data on which it is used for event selection.  The effects of overtraining on the performance of the classifiers can be mitigated by a nested cross-validation procedure, as is described in detail in Ref.~\cite{longversion}.

In using the cross validation procedure there is a danger that the bin counts become non-Poissonian due to correlations between the selections, which would need to be accounted for with computationally expensive test statistic calibration based on a large number of simulated toys. If this were found to be prohibitive in a specific application, a simple test-train split is remains a possibility to avoid this difficulty. However, we find in our tests that this does not distort the test statistic distributions in our examples in Ref.~\cite{longversion}, and we find that asymptotic formulae~\cite{Cowan:2010js} or throwing toys with counts based on the merged selected events provide accurate $p$-values.

To summarize, the extended bump hunt algorithm proceeds as follows, for a single resonance mass hypothesis $\hat{m}_\text{res}$:
\begin{enumerate}
	\item Identify an observable $m_\text{res}$ in which a signal is expected to be resonant, and a set of auxiliary variables $Y$ that are to be used for signal selection.  The variables $Y$ must be independent of $m_\text{res}$.  There are a number of methods for correcting this if not inherently true~\cite{Louppe:2016ylz,Moult:2017okx,Dolen:2016kst,Shimmin:2017mfk,Aguilar-Saavedra:2017rzt,ATL-PHYS-PUB-2018-014,Stevens:2013dya,Sirunyan:2017dnz}.  A background model $f(m_\text{res})$ is needed for $m_\text{res}$.  Typically (and in the example below) this is done with a parametric fit, though non-parametric methods are also possible~\cite{Frate:2017mai}.
	\item Define a signal region in a window around $\hat{m}_\text{res}$.
	\item Define sideband regions that are disjoint from the signal region but still sufficiently close that the background distribution in $Y$ is expected to be nearly identical.
	\item Use a cross-validation procedure to separate training samples from test samples. For each test subsample: 
	\begin{enumerate}
		\item Train a classifier to discriminate training events drawn from the sideband regions from those drawn from the signal region, using variables $Y$.
		\item Select a fraction $\epsilon$ of the most signal-like test events as determined by the classifiers.
	\end{enumerate}
	\item Merge selected event samples.
	\item Perform a statistical test for the presence of an excess in the signal region of the $m_\text{res}$ distribution after the cut has been applied, using the data outside of the signal region for background determination using the background model $f(m_\text{res})$.  The statistical analysis can be performed using pseudo-experiments generated by sub-sampling from the data itself. 
\end{enumerate}
This procedure is repeated starting from step 2 for a series of resonance mass hypotheses, as in a usual bump hunt. This entails the usual trials factor associated with the scan over the resonance variable, but does not invoke any additional trials factor associated with the space of auxiliary variables.  Using asymptotic formulae~\cite{Cowan:2010js} or throwing toys with counts based on the merged selected events provide accurate $p$-values~\cite{longversion}.

\begin{figure*}[!t]%
\centering
\subfloat{\includegraphics[width=\textwidth]{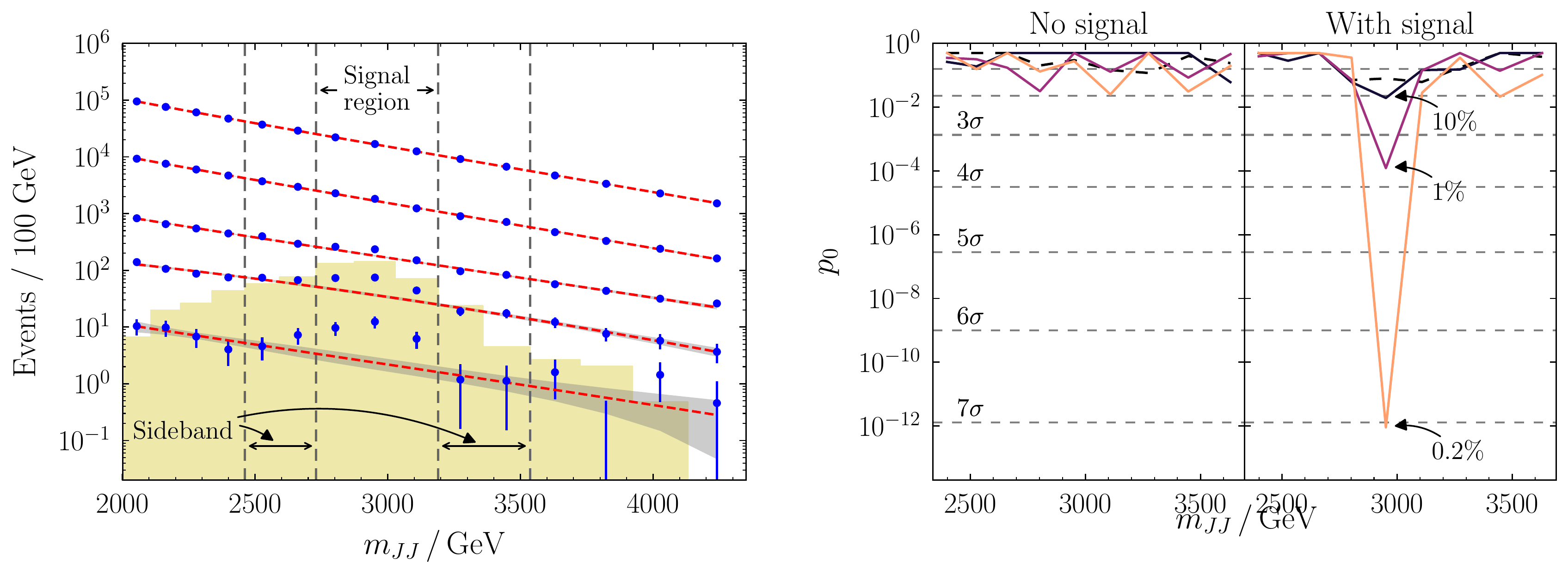}}%
\caption{\textbf{Left:} $m_{JJ}$ distribution of dijet events (including injected signal, indicated by the filled histogram) before and after applying jet substructure cuts using the NN classifier output for the $m_{JJ} \simeq 3 \; \text{TeV}$ mass hypothesis. The dashed red lines indicate the fit to the data points outside of the signal region, with the gray bands representing the fit uncertainties. The top set of markers represent the raw dijet distribution with no cut applied, while the subsequent sets of markers have cuts applied at thresholds with efficiency of $10^{-1}$, $10^{-2}$, $2\times10^{-3}$, and $2\times10^{-4}$. \textbf{Right:} Local $p_0$-values for a range of signal mass hypotheses in the case that no signal has been injected (left), and in the case that a $3 \; \text{TeV}$ resonance signal has been injected (right). The dashed lines correspond to the case where no substructure cut is applied, and the various solid lines correspond to cuts on the classifier output with efficiencies of $10^{-1}$, $10^{-2}$, and $2\times10^{-3}$.}%
\label{fig:pvalues}%
\end{figure*}

As a concrete example of the new bump hunting strategy, suppose there is a new resonance that decays into unusual jets.  We do not know a priori how to look for the new resonance, but we can consider the substructure of each jet to look for an anomalous radiation pattern.  The left plot of Fig.~\ref{fig:pvalues} shows the invariant mass of two jet four-vectors in simulated QCD dijet events~\footnote{Events are generated with Madgraph5\_aMC@NLO~\cite{Alwall:2014hca} v2.5.5 + Pythia 8.226~\cite{Sjostrand:2007gs} + Delphes 3.4.1~\cite{deFavereau:2013fsa}. For further details, see Ref.~\cite{longversion}.}.  To illustrate the power of the technique, we have also injected events from the decay of a $W'$ particle with a mass of 3 TeV.  This $W'$ is constructed to decay to a $W$ boson ($m_W\approx 80$ GeV) and a new $X$ particle ($m_X\approx 400$ GeV), which itself decays into two $W$ bosons, as described in Ref.~\cite{Agashe:2016rle,Agashe:2017wss,boosted_diboson}.   We consider the all-hadronic channel in which each $W$ boson decays into quark pairs.  The signal is thus characterized as having two large jets, one with a two-prong substructure and one with a four-prong substructure.  The shaded histogram in the left plot of Fig.~\ref{fig:pvalues} peaks at the resonance mass of 3 TeV with a broad width due to jet fragmentation and clustering effects.   Without any selection on the jets' substructure, there is no significant indication of the signal hiding under the smooth background from generic quarks and gluons.

To enhance the sensitivity of this search using the extended bump hunt method described above, a suite of classifiers are trained to distinguish a sliding signal region from sideband regions.  For each jet, the following substructure information ($Y$) is used:

\begin{equation}
m_J, \sqrt{\tau_1^{(2)}} / \tau_1^{(1)}, \tau_{21}, \tau_{32}, \tau_{43}, n_\text{trk},
\end{equation}

\noindent where $m_J$ is the jet mass, $n_\text{trk}$ is the number of charged particles (tracks) in the ungroomed jet, the $N$-subjettiness ratios are defined by $\tau_{MN} = \tau_M^{(1)} / \tau_N^{(1)}$, and the observables $\tau_N^{(\beta)}$ are defined in Ref.~\cite{Thaler:2010tr}.  

The output of the classifiers are then used to select signal-like events over the full range of the $m_{JJ}$ distribution. The resulting distributions are shown in Fig.~\ref{fig:pvalues} (left) after applying thresholds on the NN output with overall efficiencies 10\%, 1\%, 0.2\%, and 0.02\%, respectively, in descending order. Prior to applying any threshold, the resonant signal has $S/B = 6.4 \times 10^{-3}$ and significance $S/\sqrt{B} = 1.8$ in the signal region and the $m_{JJ}$ distribution has no discernible resonant feature. However, after applying the threshold determined by the classifier, a clear bump develops in the signal region with local significance of $7 \sigma$ at the 0.2\% threshold. Of course, in the event that the resonance mass is not known in advance then a scan must be performed over possible resonance masses. It is important that the procedure does not create fake bumps in the background when no signal is present. We show in  Fig.~\ref{fig:pvalues} (right) the $p$-values obtained in the mass scan over this distribution the case that (a) no signal is present, and (b) the case that the signal has been injected. We find that no significant bumps are created in the signal-free test.  Furthermore, we find that traditional searches aimed at finding di-boson resonances using jet substructure-based supervised learning algorithms (but for SM bosons) are not able to enhance the significance of this signal for a wide range of $S/B$ and classifier working points~\cite{longversion}.

\begin{figure*}%
\centering
\subfloat{\includegraphics[width=\textwidth]{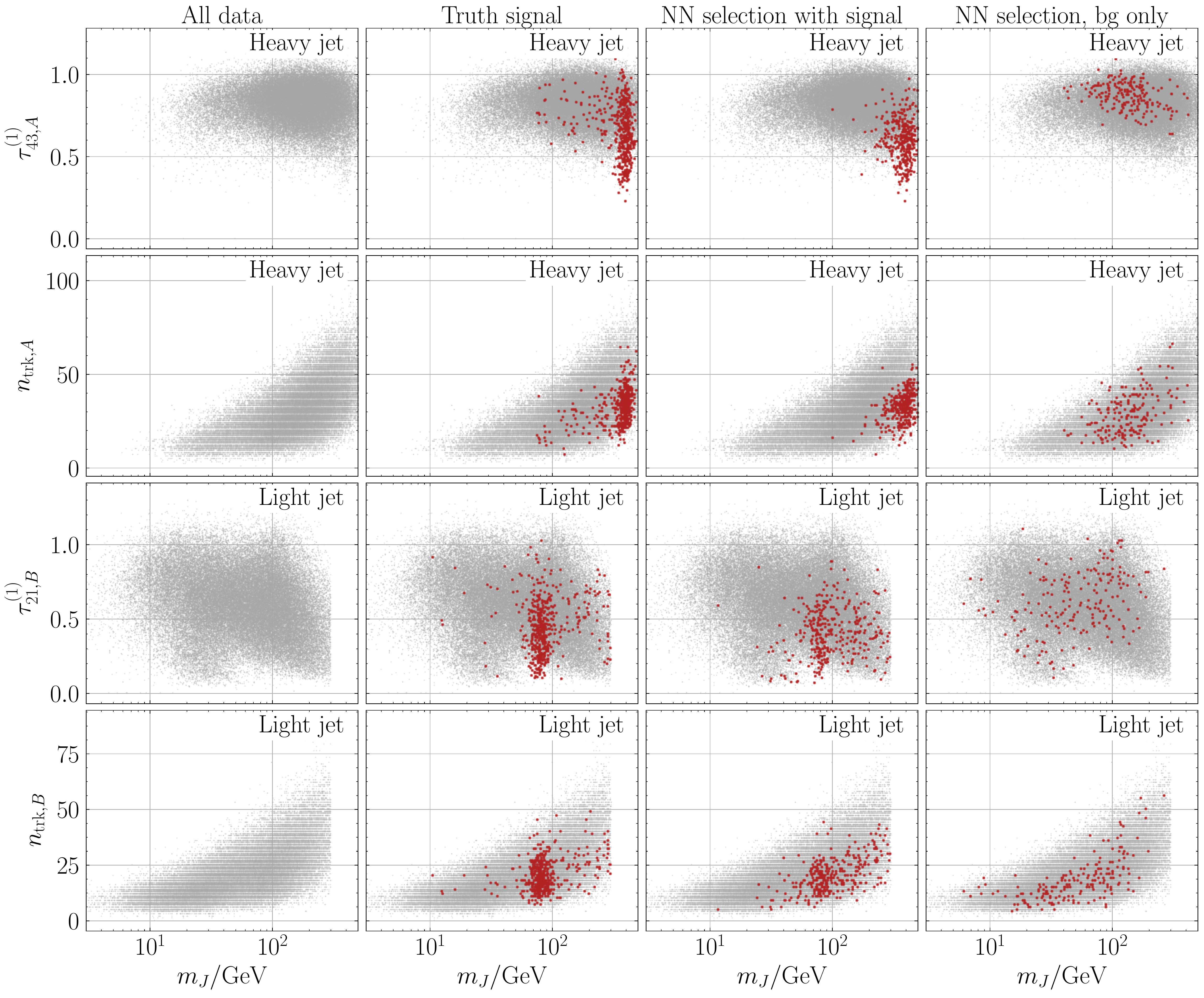}}%
\caption{2D projections of the 12D feature-space of the signal region dataset. \textbf{First column:} all signal region events.  This is repeated in the other columns to aid comparisons. \textbf{Second column:} truth-level simulated signal events highlighted in red. \textbf{Third column:} The red dots are the 0.2\% most signal-like events selected by the classifier described in the text. \textbf{Fourth column:} The red dots are the 0.2\% most signal-like events selected by a classifier trained on the same sample but with true-signal events removed.}%
\label{fig:scatterarray}%
\end{figure*}

In order to characterize the signal that the classifier has found, we can study the distribution of selected signal-like events, as illustrated in Fig.~\ref{fig:scatterarray}. We see that the classifier trained in the presence of a true signal has identified a population of events with a heavier jet with mass $m_{J \, A} \simeq 400 \; \text{GeV}$, a small number of tracks, and small $\tau_{43}$, and a lighter jet with mass $m_{J \, B}$, a small number of tracks, and small $\tau_{21}$.

In conclusion, we have presented a new technique to search for physics beyond the SM that requires very little prior knowledge of the signal.  The method was demonstrated in simulation on an all-hadronic resonance search at the LHC, where an uninteresting excess was enhanced to a level of discovery.  There are many other possibilities for applying this technique directly to data, in any case where the signal is expected to be localized in one dimension.  By naturally exploiting the power of modern machine learning, we hope that this extended bump hunt will help to expose new distance scales in nature on the quest for BSM at the LHC and beyond.

The datasets and code used for the case study can be found at Refs.~\cite{cwola_hunting_dataset, cwola_hunting_code}.

\acknowledgments
We appreciate helpful discussions with and useful feedback on the manuscript from Timothy Cohen, Aviv Cukierman, Patrick Fox, Jack Kearney, Zhen Liu, Eric Metodiev, Brian Nord, Bryan Ostdiek, Francesco Rubbo, and Jesse Thaler. We would also like to thank Peizhi Du for providing the UFO file for the benchmark signal model. The work of JHC is supported by NSF under Grant No. PHY-1620074 and by the Maryland Center for Fundamental Physics (MCFP). The work of B.N. is supported by the DOE under contract DE-AC02-05CH11231.
This manuscript has been authored by Fermi Research Alliance, LLC under
Contract No.  DE-AC02-07CH11359 with the U.S. Department of Energy,
Office of Science, Office of High Energy Physics. The United States
Government retains and the publisher, by accepting the article for
publication, acknowledges that the United States Government retains a
non-exclusive, paid-up, irrevocable, world-wide license to publish or
reproduce the published form of this manuscript, or allow others to do
so, for United States Government purposes.

\appendix

%\section{Simulation Details}
%\label{app:sim}

%\section{Classifier Details}
%\label{app:class}

\bibliography{myrefs}

\end{document}